\documentclass[aps,prl,twocolumn,groupedaddress,english]{revtex4-1}
\usepackage[utf8]{inputenc}
\usepackage[dvips]{graphicx}
\usepackage{epsfig}
\usepackage[fleqn]{amsmath}
\usepackage{color}
\usepackage{float}

\begin{document}
\title{Ion-molecule reactions below 1~K: Observation of a strong enhancement of the reaction rate of the ion-dipole reaction $\mathrm{He}^++\mathrm{CH}_3\mathrm{F}$}

\author{Valentina Zhelyazkova}
\author{Fernanda B. V. Martins}
\author{Josef A. Agner}
\author{Hansj\"urg Schmutz}
\author{Fr\'ed\'eric Merkt}
\email[]{merkt@phys.chem.ethz.ch}
\affiliation{Laboratory of Physical Chemistry, ETH Zurich, CH-8093 Zurich, Switzerland}

\date{\today}

\begin{abstract}
The reaction between He$^+$ and CH$_3$F forming predominantly CH$_2^+$ and CHF$^+$ has been studied at collision energies $E_{\rm coll}$ between 0 and $k_{\rm B}\cdot 10$~K in a merged-beam apparatus. To avoid heating of the ions by stray electric fields, the reaction was observed within the orbit of a highly excited Rydberg electron. Supersonic beams of CH$_3$F and He($n$) Rydberg atoms with principal quantum number $n= 30$ and 35 were merged and their relative velocity tuned using a Rydberg-Stark decelerator and deflector, allowing an energy resolution of 150 mK. A strong enhancement of the reaction rate was observed below $E_{\rm coll}/k_{\rm B} = 1$~K. The experimental results are interpreted with an adiabatic capture model that accounts for the state-dependent orientation of the polar CH$_3$F molecules by the Stark effect as they approach the He$^+$ ion. The enhancement of the reaction rate at low collision energies is primarily attributed to para-CH$_3$F molecules in the $J=1,\,KM=1$ high-field-seeking states, which represent about 8\% of the population at the 6~K rotational temperature of the supersonic beam.

\end{abstract}

\maketitle
Considerable progress has recently been made in experimental studies of molecular collisions and reactions at low temperatures and collision energies (see, e.g., Refs. \cite{gilijamse06a,bell09a,ni10a,parazzoli11a,willitsch12a,henson12a,chefdeville13a,stuhl15a,jankunas14a,heazelwood15a,perreault17a,dejongh20a} and references therein). Investigations below 1~K have become possible for processes involving neutral molecules \cite{ni10a,henson12a,jankunas14a,perreault17a,zou19a,bibelnik19a,gawlas20a}. Less progress has been made for ion-molecule reactions, primarily because ions are easily heated by stray fields, which makes the low temperature range very difficult to reach. Methods relying on supersonic beams from Laval nozzles~\cite{marquette85a} and guided ion beams and traps~\cite{gerlich08a,paetow10a,tran18a,markus20a} have proven successful to study ion-molecule reactions down to about 10~K.
Approaches to reach the range below 1~K relying on ion traps and Coulomb crystals are being developed and promising results were reported \cite{bell09a,willitsch12a,doerfler19a}. In the absence of experimental data on ion-molecule reactions below 10~K, theory has challenged experiments for several decades \cite{clary90a}, and it is only today that experiment can respond.

In this letter we present, with the example of the reaction between He$^+$ and CH$_3$F, experimental results at very low collision energies on barrier-free exothermic reactions between ions and polar molecules, which are the class where the strongest low-temperature effects are expected \cite{ausloos78a,bowers79a,ng92b,clary85a,troe87a}. We report the observation of a large enhancement of the reaction rate at collision energies $E_{\rm coll}$ below $E_{\rm coll}/k_{\rm B}$ below 1~K related to the gradual orientation, upon approach of the ion, of the polar molecule by the Coulomb field of the ion. Numerous models and approximations have been used to predict reaction rate constants of ion-dipole reactions below 10 K over the past 50 years \cite{ausloos78a,bowers79a,ng92b,clary85a,clary85b,clary90a,stoecklin92a,troe87a,troe96a,auzinsh13a,auzinsh13b,dashevskaya05a,dashevskaya16a}. These models typically predict enhancements of the reaction rates at the lowest temperatures and collision energies, however, with large differences concerning the extent of the enhancement and the range over which it takes place. Our new results provides a basis for the assessment of these models.

Experimentally we have extended a method originally developed to study the reaction H$_2^+$ + H$_2\rightarrow$ H$_3^+$ + H at collision energies below 1~K \cite{allmendinger16a,allmendinger16b} to the reactions of He$^+$ with polyatomic molecules. In brief, the ion reactant (He$^+$) is substituted by the corresponding Rydberg system (He$(n)$) in a state of high principal quantum number $n$. The Rydberg electron acts as a distant spectator, does not affect the reaction \cite{pratt94a,wrede05a,matsuzawa10a}, and serves the purpose of shielding the ions from stray fields. The method combines the advantages of merged-beams approaches to study neutral-neutral reactions at low collision energies \cite{henson12a,jankunas14a,perreault17a} with those offered by Rydberg atoms and molecules, e.g., the ease of manipulation of their translational motion with inhomogeneous electric fields \cite{procter03a,vliegen04a,merkt19a}.
\begin{figure*}
	\includegraphics[width = 1.0\textwidth,trim={0.5cm 1cm 1cm 0},clip]{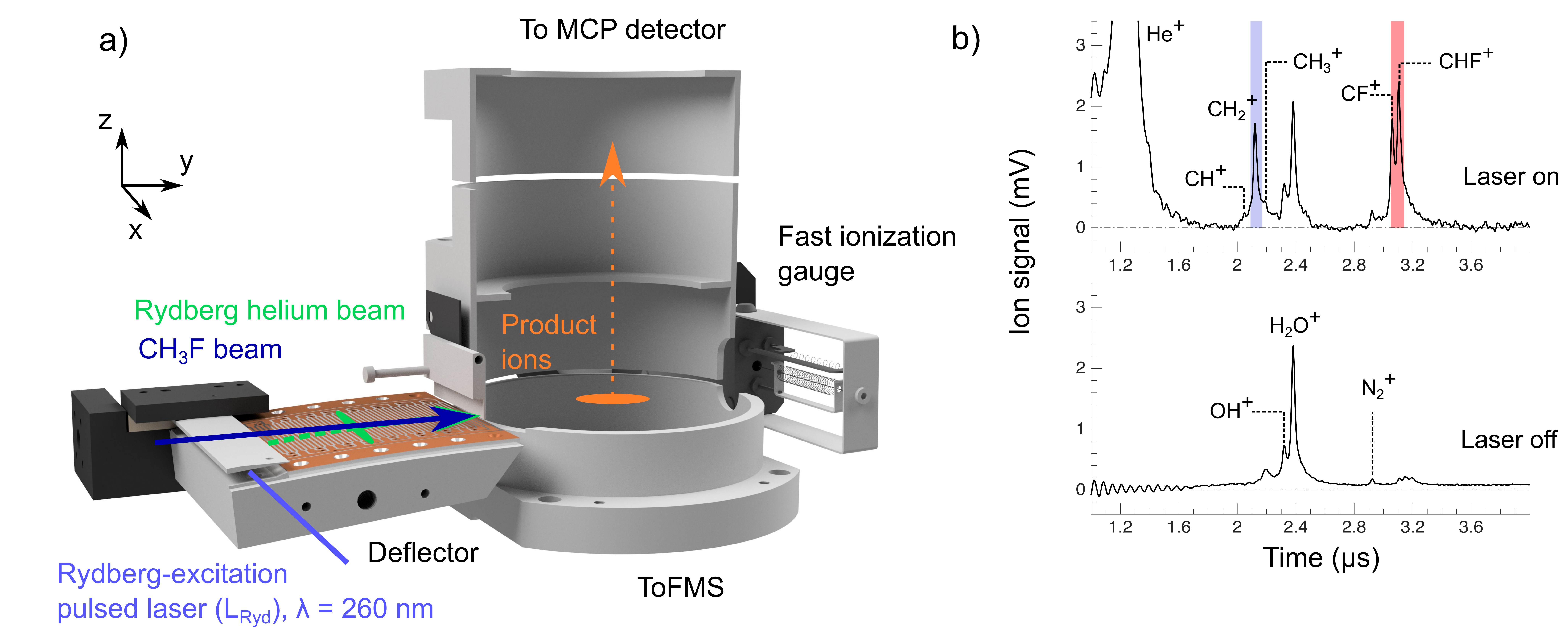}
	\caption{\label{fig1} a) Schematic view of the merged-beam setup used to study the $\mathrm{He}^++\mathrm{CH}_3\mathrm{F}$ reaction at low collisional energies, with the 50-electrode Rydberg-Stark deflector and the ToFMS. b) Examples of ToF mass spectra of the product ions recorded with (top panel) and without (bottom panel) L$_{\mathrm{Ryd}}$. The blue and red areas indicate the regions used to monitor the relative reaction yields (see text).}
\end{figure*}
\newline
\indent
The CH$_3$F and He beams are produced in supersonic expansions of pure CH$_3$F and He gas, respectively, through the orifices of two home-made pulsed valves (pulse duration $\leq 20$~$\mu$s, repetition rate 25~Hz). The two beams initially propagate along axes separated by a 5$^{\circ}$ angle. The helium atoms are excited to a Rydberg-Stark state and subsequently merged with the ground-state CH$_3$F beam using a surface-electrode Rydberg-Stark
deflector~\cite{allmendinger14a,allmendinger16a,zhelyazkova19a}. After the CH$_3$F and Rydberg-He beams are merged, they enter a time-of-flight mass spectrometer (ToFMS) (Fig.~\ref{fig1}a), where the reaction product ions are extracted and detected with a microchannel plate (MCP) detector. The center of the ToFMS along the merged-beam propagation axis is defined as $y=0$. 

The CH$_3$F valve, positioned at $y\simeq-62$~cm, is temperature-stabilized to $330\pm0.4$~K, corresponding to a mean forward velocity of the beam of $\sim870$~m/s. We use two identical fast-ionization gauges separated by 20 cm to measure the velocity distribution of the CH$_3$F beam. The CH$_3$F beam is skimmed twice to select its central, coldest part before it is merged with the deflected Rydberg helium sample (referred to as He($n$) henceforth). The relative populations of the rotational levels of the ortho- and para-CH$_3$F molecules in the beam both correspond to a temperature $T_{\rm rot}$ of $\approx 6$~K, with only seven $J_K$ states significantly populated at 6 K, i.e., $0_0$ (9.5\%), $1_0$ (19.0\%), $2_0$ (13.9\%), $3_0$ (5.8\%), $1_1$ (23.5\%), $2_1$ (17.3\%) and $3_1$ (7.1\%). Ortho-para conversion does not take place in the supersonic expansion and the beam consists of 50\% of the para ($K=1$ states) and 50\% of the ortho ($K=0$ states) nuclear-spin isomer \cite{hanson16a}.

The helium valve is positioned at $y\simeq-100$~cm, and slightly displaced below the CH$_3$F beam-propagation axis. A pulsed electric discharge at the valve orifice populates the metastable (1s)$^1$(2s)$^1$ $^3$S$_1$ state of helium (referred to as He* below)~\cite{halfmann00a,zhelyazkova19a}. The copper body of the valve is cooled with a two-stage pulse-tube cryo-cooler and temperature-stabilized to $100\pm0.1$~K. The mean velocity of the He* beam in the forward direction is $\sim 1040$~m/s.
The He* beam passes through two skimmers before it intersects a pulsed-laser beam of wavelength $\sim260$~nm (L$_{\mathrm{Ryd}}$ in Fig.~\ref{fig1}a) in the presence of an electric field \cite{zhelyazkova19a}. The He* atoms are excited by the laser to a low-field-seeking $n$ Rydberg state labeled by $k$ ($k=-(n - |m_{\ell}| -1):2:(n - |m_{\ell}|-1)$, where $m_\ell$ is the magnetic quantum number \cite{gallagher94a}). The laser radiation is linearly polarized parallel to the electric field and thus $m_{\ell}=0$. For the experiments presented here, $(n,k) = (30,25)$ or $(35,26)$. After excitation, the He($n$) atoms fly over the deflector, which consists of an array of 50 electrodes (dimensions 1~mm~$\times$~30~mm in the $y$ and $x$ dimensions, respectively), attached to a curved substrate. 

The design and operation principle of the deflector are described in Refs.~\cite{allmendinger14a,zhelyazkova19a}. The defelctor can also be used to accelerate or decelerate the He(n) beam by applying chirped oscillatory potentials of amplitude $V_{\rm amp}$ to the electrodes \cite{hogan12b}.

After the He($n$) atoms are merged with the CH$_3$F beam, they enter the ToFMS. To define the 7-$\mu$s-long time interval during which the reaction is monitored and to optimize the energy resolution of the experiment \cite{allmendinger16a}, we apply a sequence of two 3-$\mu$s-long electric-field pulses separated by a field-free interval of 7~$\mu$s. The first pulse sweeps all ions produced prior to the selected reaction-time interval out of the reaction zone. The second pulse then extracts all ions generated during the selected 7-$\mu$s-long reaction time toward the MCP detector. 
Examples of ToF traces of the ion products of the $\mathrm{He}(n)+\mathrm{CH}_3\mathrm{F}$ and $\mathrm{He}^*+\mathrm{CH}_3\mathrm{F}$ reactions are displayed in Fig.~\ref{fig1}b. In the lower trace, obtained without L$_{\mathrm{Ryd}}$, a prominent peak corresponding to H$_2$O$^+$ is observed, in addition to several smaller peaks assigned to OH$^+$, N$_2^+$, CH$_{p\in \{1,2,3\}}^+$ and CH$_{q\in \{0,1,2\}}{\rm F}^+$ ions. These ions originate from Penning ionization processes between the He* atoms and either background water, nitrogen, or CH$_3$F molecules in the vacuum chamber. When L$_{\mathrm{Ryd}}$ is turned on, several additional peaks become visible (upper trace). The first broad peak, at arrival times $<1.6$~$\mu$s, originates from field-ionized He($n$) atoms. The other peaks correspond to the CH$^+$, CH$_2^+$, CH$_3^+$, CF$^+$ and CHF$^+$ products of the He($n$)+CH$_3$F reaction and indicate a rich short-range chemistry. 

The dominant product ions are CH$_2^+$, CF$^+$ and CHF$^+$. We determine the relative reaction yield from the integrated signals within the blue (I$_{\mathrm{CH}_2^+}$) and red (I$_{\mathrm{CF}^++\mathrm{CHF}^+}$) integration windows indicated in the top panel of Fig.~\ref{fig1}b and monitor them as a function of $E_{\mathrm{coll}}\ (= \mu v_{\mathrm{rel}}^2/2)$, given by the selected relative velocity $v_{\mathrm{rel}}$ of the two beams ($\mu$ is the reduced mass of the He$^+$ + CH$_3$F collision partners).

As in previous merged-beam studies of neutral-neutral reactions at low collision energies \cite{henson12a,jankunas15a}, we exploit the velocity dispersion resulting from the short opening time of the CH$_3$F valve ($\leq 20\,\mu$s) and the long distance between the valve orifice and the reaction region to select a very narrow velocity range ($\approx \pm 5$~m/s) of the CH$_3$F beam around the chosen velocity of 990~m/s. 
By applying the appropriate potential waveforms to the deflector, we scan the He($n$)-beam velocity $v_{\mathrm{He(\textit{n})}}$ in the 750-1200~m/s range, corresponding to relative velocities $v_{\mathrm{rel}} = v_{\mathrm{CH_3F}}-v_{\mathrm{He(\textit{n})}}$ between $240$~m/s and $-210$~m/s
and collisions energies $E_{\mathrm{coll}}$ between 0 and $(12\ {\rm K})\cdot k_{\mathrm{B}}$. 

\begin{figure}
\includegraphics [width=1\columnwidth,trim={0.1cm 0.1cm 0cm 0},clip] {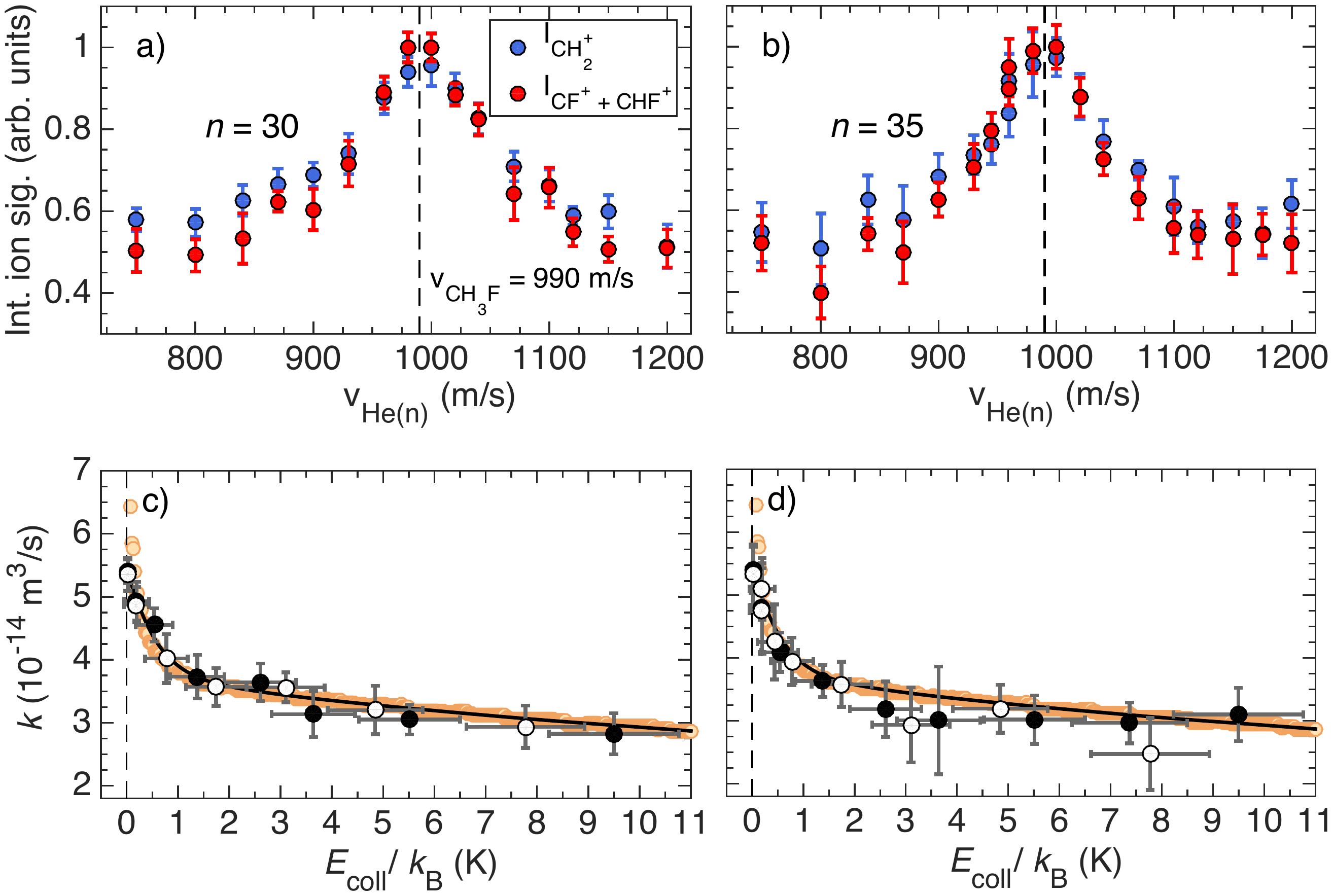}	
	\caption{\label{fig2} a) and b): Measured normalized signals corresponding to the CH$_2^+$ ($I_{\mathrm{CH}_2^+}$, blue circles) and CF$^+$ + CHF$^+$ ($I_{\mathrm{CF}^+ + \mathrm{CHF}^+}$, red circles) product ions of the reaction He$^+$+CH$_3$F as a function of the selected final longitudinal velocity $v_{\mathrm{He}(n)}$ of the beam of He atoms excited to $(n,k) = (30,25)$ (a) and $(35,26)$ (b) Rydberg-Stark states. The dashed vertical lines indicate the selected CH$_3$F-beam longitudinal velocity. c) and d): Comparison of the integrated total product-ion signals ($I_{\mathrm{tot}}$ =
		$I_{\mathrm{CH}_2^+}$ + $I_{\mathrm{CF}^++\mathrm{CHF}^+}$) for $v_{\mathrm{He}(n)} >v_{\mathrm{CH_3F}}$ (black dots) and $v_{\mathrm{He}(n)} <v_{\mathrm{CH_3F}}$ (open circles) as a function of $E_{\mathrm{coll}}/k_{\mathrm{B}}$ with the calculated reaction rate constants averaged over the CH$_3$F rotational states for $T_{\mathrm{rot}}=6$~K, with (black line) and without (yellow circles) the effect of the average over the distribution of collision energies.}
\end{figure}
Decelerating or accelerating the He($n$) atoms away from their initial velocity of 1040~m/s leads to losses. To account for this effect and Rydberg-atom-lifetime effects, the values of $I_{\mathrm{CH}_2^+}$ and $I_{\mathrm{CF}^++\mathrm{CHF}^+}$ measured at each value of $v_{\mathrm{He(\textit{n})}}$ are normalized to the total amount of He($n$) that reaches the center of the ToFMS. This normalization factor is determined in separate pulsed-field-ionization measurements carried out by applying a pulsed potential of 5~kV to the bottom electrode of the ToFMS and detecting the resulting He$^+$ ions on the MCP. 
By scanning the time at which the field-ionization pulse is applied, we also determine the time $t_{c,v_{\mathrm{He(\textit{n})}}}$ at which the center of the Rydberg-atom packet reaches $y = 0$ with respect to the time of laser excitation. The CH$_3$F-valve opening time is then chosen such that the molecules traveling at 990~m/s also arrive at position $y = 0$ at time $t_{c,v_{\mathrm{He(\textit{n})}}}$, which corresponds to the middle of the 7-$\mu$s-long time interval during which the reaction is monitored.

Displayed in the top panels of Fig.~\ref{fig2} are the normalized ion signals $I_{\mathrm{CH}_2^+}$ and $I_{\mathrm{CF}^++\mathrm{CHF}^+}$ as a function of $v_{\mathrm{He}(n)}$ obtained from reactions with He($n$) atoms excited to $(n,k)=(30,25)$ (a) and $(35,26)$ (b) Rydberg-Stark states. The data points and associated vertical error bars correspond to averages and standard deviations, respectively, each determined from 8000 experimental cycles. To be able to directly compare the data recorded at the two different values of $n$, $V_{\rm amp}$ was chosen such that the depths of the moving traps were the same for both sets of Rydberg states. The loading of the Rydberg-atom beam into the moving traps was carefully optimized to avoid heating, resulting in a translational temperature of $T_{\mathrm{He}(n)}\approx 140(60)$~mK at both $n$ values, as estimated from particle-trajectory simulations.

The measured values of both $I_{\mathrm{CH}_2^+}$ and $I_{\mathrm{CF}^++\mathrm{CHF}^+}$ depicted in Fig.~\ref{fig2}a,b increase and reach a maximum at $v_{\mathrm{He(\textit{n})}}=v_{\mathrm{CH_3F}}=990$~m/s ($v_{\mathrm{rel}}=0$), which corresponds to approximately twice the signal strengths measured at $v_{\mathrm{He(\textit{n})}}=750$~m/s and 1200~m/s. 
The collision-energy dependence of the total product-ion yield $I_{\mathrm{tot}} =
I_{\mathrm{CH}_2^+} + I_{\mathrm{CF}^++\mathrm{CHF}^+}$ for $v_{\mathrm{rel}}<0$ (black dots) and $v_{\mathrm{rel}}>0$ (open circles) is presented in Fig.~\ref{fig2}c,d. It reveals a slow increase as $E_{\mathrm{coll}}/k_{\mathrm{B}}$ decreases from 10 to 1 K, and a sharp increase below 1 K. 
The data sets obtained at $n=30$ and 35 confirm the expectation that the Rydberg electron merely acts as a spectator in fast capture-type reactions, as already demonstrated in our previous studies of the H$_2$ +  H$_2^+$ reaction \cite{allmendinger16a,allmendinger16b}.

The gray horizontal bars in Fig.~\ref{fig2}c,d represent the ranges of collision energies $\Delta E_{\rm coll}$ probed experimentally at the selected values of $E_{\rm coll}$, and correspond to the energy resolution of our measurements, given by
\begin{eqnarray}
\frac{\Delta E_{\mathrm{coll}}}{k_{\mathrm{B}}} &=& \Delta T_{\mathrm{res}} + 2\sqrt{\Delta T_{\mathrm{res}}}\sqrt{\frac{E_{\mathrm{coll}}}{k_{\mathrm{B}}}}.
\label{eq1}
\end{eqnarray}
In Eq.~(\ref{eq1}), $\Delta T_{\mathrm{res}}=150$~mK corresponds to the estimated distribution of relative velocities when $E_{\mathrm{coll}}$ is zero. It is obtained by adding in quadrature the translational temperatures of the two beams ($T_{{\rm He}(n)}\approx 140$~mK, and $T_{{\rm CH}_3{\rm F}}\approx50$~mK) in the reaction zone, as determined from numerical particle-trajectory simulations in the case of the He($n$) beam and from measurements with the fast-ionization gauges in the case of CH$_3$F. 
 
\begin{figure}
	\centering
\includegraphics[width=1.2\columnwidth,trim={1.9cm 0.25cm 0cm 0},clip]{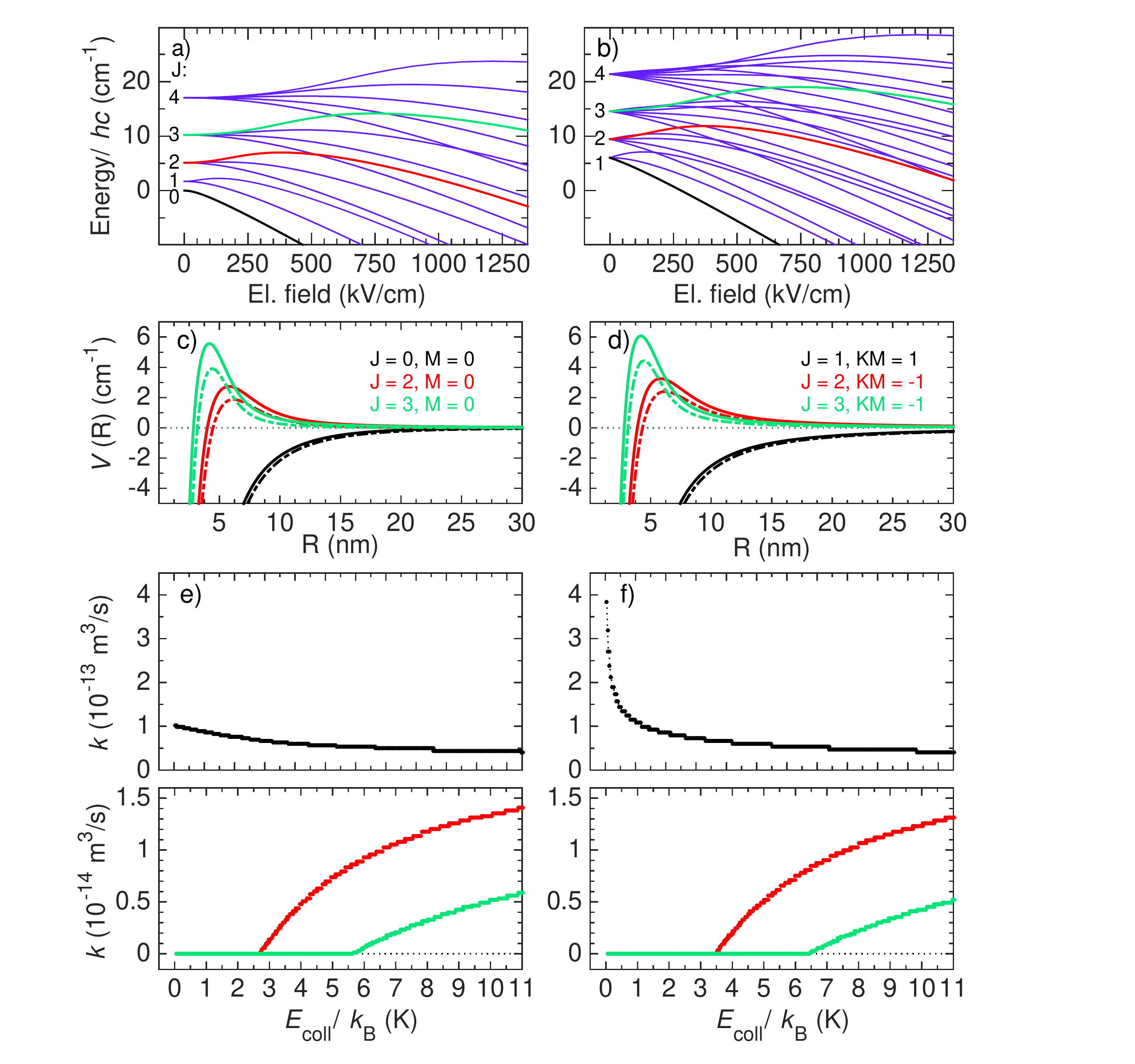}
	%[width = 0.8\textwidth]
	\caption{\label{fig3} a) and b): Calculated energies in CH$_3$F for states with $K = 0$ (a) and $|K| = 1$ (b) in the presence of an electric field. (c) and (d): Adiabatic interaction potentials for the $\mathrm{He}^++\mathrm{CH_3F}(J,K,M)$ reaction, $V(R)$, for CH$_3$F in the states highlighted in black, red and green in a) and b), for $\ell = 0$ (dash-dotted lines) and $\ell=25$ (solid lines) (see text for details). (e) and (f): Calculated values of the reaction rate constants for CH$_3$F in the indicated states. a), c) and e): $K=0$; b), d) and f): $|K|=1$.}
\end{figure}

 CH$_3$F has a polarizability volume $\alpha^\prime$ of $2.54\times 10^{-30}$~m$^3$ \cite{olney97a}, corresponding to a Langevin-capture rate constant $k_{\rm L}=\sqrt{\pi\alpha^\prime e^2/(\mu\epsilon_0)}$ of $1.97\times{}10^{-15}$~m$^3$/s. 
 It has a permanent electric dipole moment $\mu_{\rm el}$ of $6.20\times 10^{-30}$~Cm \cite{marshall80a}.
 In one of the standard capture models for ion-dipole reactions, based on the locked-dipole approximation (LDA) \cite{bowers79a}, the dipole moment is assumed to be locked with the energetically favorable orientation along the collision axis, which results in the long-range interaction potential 
\begin{equation}\label{FDA}
V_{{\rm LDA},\ell}(R)= -\frac{\alpha^\prime e^2}{8\pi\epsilon_0R^4} - \frac{e\mu_{\rm el}}{4\pi\epsilon_0R^2} + \frac{\hbar^2\ell(\ell+1)}{2\mu R^2}.
\end{equation}
The LDA leads to an expression for the energy-dependent rate constant 
\begin{equation}\label{rate_FDA}
k_{\rm LDA}(E_{\rm coll})= k_{\rm L} + \frac{e\mu_{\rm el}}{2^{3/2}\epsilon_0\sqrt{\mu E_{\rm coll}} }
\end{equation}
that diverges at $E_{\rm coll}=0$. The corresponding thermal rate constant 
\begin{equation}\label{rate_FDA_T}
k_{\rm LDA}(T)= k_{\rm L} + \frac{e\mu_{\rm el}}{\epsilon_0} \sqrt{\frac{1}{2\pi\mu k_{\rm B}T}}
\end{equation}
is $4.03\times 10^{-13}$~m$^3$/s at the effective temperature of 150~mK in our experiments at $E_{\rm coll}=0$. The LDA model predicts enhancement factors $k_{\rm LDA}(T=150\ {\rm mK})/k_{\rm LDA}(E_{\rm coll}=k_{\rm B}\cdot 10\ {\rm K})$ and  $k_{\rm LDA}(T=150\ {\rm mK})/k_{\rm LDA}(E_{\rm coll}=k_{\rm B}\cdot 4\ {\rm K})$ of 8.86 and 5.49, respectively, whereas our experiments yield values of 1.9 and 1.6 for these factors.

To better describe the rate constants of ion-dipole reactions at low collision energies, we found it necessary to consider the initial distribution of the polar molecules over the rotational levels $|JKM\rangle$ in the supersonic beam and how these states are affected by the growing electric-field strength they experience upon approach of the He$^+$ ion \cite{clary85a,troe87a}. To analyze our experimental results and test more advanced models, we adapt the statistical adiabatic-channel model for ion--dipole collisions of Troe \cite{troe87a} to our experimental situation. We consider that the signal observed at each value of $E_{\rm coll}$ in Fig.~\ref{fig2}c,d is proportional to a reaction rate $\langle k(\bar{E}_{\rm coll})\rangle$ that corresponds to the average over a Gaussian distribution $G(E_{\rm coll})$ of collision energies centered at the selected $E_{\rm coll}$ value, with widths given by Eq.~(\ref{eq1}):
\begin{equation}\label{rate_sum_sum}
\langle k(\bar{E}_{\rm coll})\rangle=\int_0^\infty G(E_{\rm coll})k(E_{\rm coll}){\rm d}E_{\rm coll}.
\end{equation}

The values of $k(E_{\rm coll})$ are determined as sums of state-specific rate coefficients over all CH$_3$F $j=|JKM\rangle$ states populated at $T_{\rm rot}=6$~K:
\begin{equation}\label{rate_state_sum}
k(E_{\rm coll})=\sum_j {\rm e}^{-E_j/(k_{\rm B}T_{\rm rot})}k_j(E_{\rm coll}),
\end{equation}
where $E_j$ is the field-free energy of state $j$. Only $K=0$ (ortho) and $|K|=1$ (para) CH$_3$F rotational levels with $J\leq 4$ are significantly populated at 6~K, resulting in 25 $j=|J0M\rangle$ and 48 $j=|J\pm 1M\rangle$ states to be considered in the sum.
We then calculate the $R$-dependent adiabatic-channel potentials by adding the Stark shifts $\Delta E^{\rm Stark}_j(F)$ of the rotational levels to the Langevin-type potentials
\begin{equation}\label{ACM}
V_{j,\ell}(R)= -\frac{\alpha^\prime e^2}{8\pi\epsilon_0R^4} + \frac{\hbar^2\ell(\ell+1)}{2\mu R^2} + \Delta E^{\rm Stark}_j(F(R))
\end{equation}
up to electric fields $F$ of $4\times 10^3$~kV/cm, corresponding to a He$^+$--\,CH$_3$F distance $R$  of 1.9~nm. The calculation of Stark shifts is performed by diagonalizing the Hamiltonian matrix including the Stark effect in a basis of symmetric-top rotational wavefunctions as explained, e.g., in Ref.~\cite{meerakker12a}. The state-specific rate coefficients at a given $E_{\rm coll}$ value are obtained from the maximal $\ell$ partial wave satisfying the Langevin-capture condition $E_{\rm coll}\geq V_{j,\ell}(R)$.

The essential steps of the calculations are illustrated in Fig.~\ref{fig3}, which shows the results for $K=0$ and $\pm 1$ in the left and right columns, respectively. The top panels depict the rotational energies including Stark shifts, the middle panels show the adiabatic potentials for $\ell=0$ and 25 for three representative states, and the bottom panels the corresponding state-specific rate coefficients. Levels with a positive Stark shift (negative value of the product $KM$) exhibit long-range energy barriers and vanishing rate constants at the lowest collision energies. In contrast, the rate constants of states with negative Stark shifts at low fields continuously grow with decreasing $E_{\rm coll}$ values, the effect being particularly strong for the states subject to a linear Stark shift at low fields, as exemplified by the $J=1,KM=1$ Stark states (Fig.~\ref{fig3}f).
The yellow dots in Fig.~\ref{fig2}c,d represent the calculated state-averaged rate constants $k(E_{\rm coll})$ [Eq.~(\ref{rate_state_sum})] and the black line the rate constants $k(\bar{E}_{\rm coll})$ integrated over the range of collision energies probed at each selected collision energy [Eq.~(\ref{rate_sum_sum})]. The enhancement of the rate constants below 1~K is dominated by the contributions from the $J=1,KM=1$ states, which represent only 8\% of the total population at 6K. The calculated and experimental relative rate constants agree within the experimental error bars over the entire range of collision energies, which enables us to plot the experimental result on an absolute scale in Fig.~\ref{fig2}c,d, with a maximal value of $5.4\times 10^{-14}$~m$^3$/s at $E_{\rm coll}=0$, more than 20 times larger than $k_{\rm L}$ but 7.5 times smaller than $k_{\rm LDA}(T=150$~mK). Inspection of Fig.~\ref{fig2}a,b reveals that the enhancement of the yield of F-containing ions near $E_{\rm coll}=0$ is stronger than that of CH$_n^+$ ions. We attribute this observation to the orientation of CH$_3$F molecules in the $J=1,KM=1$ state, which exposes the F side of the molecule to the He$^+$ ion.

The experimental results presented in this letter are the first to reveal a strong enhancement of the rate of an ion-dipole reaction below 1~K. The theoretical analysis of the results leads to the determination of energy-dependent, state-specific rate coefficients, from which thermal rate constants can be calculated, and enables us to conclude that the sharp enhancement of the rates below 1 K originates from states exhibiting a negative linear Stark shift at low fields. Our results demonstrate the failure of the LDA and the validity of the adiabatic channel model in the range of $E_{\rm coll}/k_{\rm B}$ values between 0.1 and 10 K. The enhancement observed experimentally would be much larger if the CH$_3$F sample could be prepared selectively in states with positive $KM$ values and this aspect represents an attractive prospect for future work.

\begin{acknowledgments}
We thank Dr. Matija {\v{Z}}e{\v{s}}ko for help with the development of the experimental apparatus, Jo\"el Jenny for his contributions to preliminary measurements, and Katharina H\"oveler, Dr. Paul Jansen, and Dr. Johannes Deiglmayr for fruitful discussions. This work is supported by the Swiss National Science Foundation (Grant No. 200020-172620) and the European Research Council (ERC) under the European Union’s Horizon 2020 research and innovation programme (Advanced Grant No. 743121).
\end{acknowledgments}

%\bibliography{bib-file-vaze-20200526,ERC2016,grpbib,antrag,bib_SNF2016}
%

\end{document}